\documentclass[9pt,twocolumn]{article}

\usepackage{authblk}
\usepackage{graphicx}
\usepackage{amsmath}
\usepackage[english]{babel}
\usepackage{url}
\usepackage[colorlinks=true, allcolors=blue]{hyperref}

\usepackage[left=48pt,%
right=42pt,%
top=46pt,%
bottom=60pt,%
headheight=15pt,%
headsep=10pt,%
letterpaper,twoside]{geometry}%
\usepackage[labelfont={bf,sf},%
labelsep=period,%
figurename=Figure,%
singlelinecheck=off,%
justification=RaggedRight]{caption}
\usepackage[numbers,sort&compress]{natbib}
\title{Evidence of Linear Chirp in Mid-Infrared Quantum Cascade Lasers}

\author[1*]{Matthew Singleton}
\author[1]{Pierre Jouy}
\author[1]{Mattias Beck}
\author[1$\dagger$]{J\'er\^ome Faist}

\affil[1]{Institute for Quantum Electronics, ETH Zurich, 8093 Z\"urich, Switzerland}
\affil[*]{Corresponding author: smatthew@phys.ethz.ch}
\affil[$\dagger$]{e-mail: jfaist@phys.ethz.ch}

\date{}

\begin{document}
	
	\maketitle
	
	\begin{abstract}
		We measure the inter-modal phase relation of a quantum cascade laser frequency comb operating at 8$\mu$m using a coherent beatnote spectroscopy. We find these phases to be reproducible after cycling the power of the device, and to be smoothly varying with driving current. Moreover, these phases describe a comb state exhibiting a simple, linear chirp, which in fact corresponds to the lowest state of chirp to minimise the amplitude modulation, as required for combs driven by four wave mixing in a gain medium with a short gain recovery lifetime.
	\end{abstract}
	
	\section{Introduction}

	Standard monolithic Fabry-P\'erot  quantum cascade lasers (QCLs) can operate in an optical frequency comb regime in the mid-IR~\cite{hugi_mid-infrared_2012} and the terahertz~\cite{burghoff_evaluating_2015,rosch_octave-spanning_2015}, provided that their cavity exhibits a low enough group velocity dispersion (GVD). Their comb nature has been rigorously proven through intermode beat spectroscopy~\cite{hugi_mid-infrared_2012}, and later through a dual-comb frequency counting experiment \cite{villares_dual-comb_2014}. Such combs are known to arise through a cascaded four-wave mixing process\cite{hugi_mid-infrared_2012}, which acts to injection lock the residually dispersed cavity modes. These lasers do not feature singular, short pulses at the cavity round trip time, typical of other mode-locked lasers. Instead, numerical work \cite{khurgin_coherent_2014,villares_quantum_2015} suggests a near-constant temporal envelope and rather a more complicated periodic modulation in frequency. The frequency modulated nature of the comb was also suggested by experiments where frequency modulation to amplitude modulation (FM to AM) conversion was performed using an optical discriminator formed by an absorption line~\cite{hugi_mid-infrared_2012} or an unbalanced interferometer (see Figure~\ref{fig:setup}~(b)). This character arises naturally from the short sub-ps upper state lifetime, which prevents the storage of energy between cavity round trips and as such acts to dampen rather than promote pulses. 

	Key to understanding time domain behaviour, the intensity $I(t)$ and the instantaneous frequency $f_i(t)$, are the spectral phases. Initial theoretical investigations ~\cite{khurgin_coherent_2014,henry_pseudorandom_2017} predicted that the output field would be characterised by periodic rapid, quasi-random swings of the instantaneous frequency as a function of time. The complexity of such states hinted that the final lasing state would somehow be randomly chosen among any number of equally favourable states each time the laser was switched on, as was found to be the case of for some parameters in microresonator combs~\cite{delhaye_self-injection_2014}. Nonetheless, with careful control of parameters, they also observe stable, deterministic states in such systems\cite{delhaye_phase_2015}.	
	
	Owing to the largely FM nature of the output, conventional techniques~\cite{trebino_measuring_1997,walmsley_characterization_1996} based on non-linear conversion of the optical pulses cannot be applied to characterise the field. For this reason, the initial comb operation of quantum cascade laser was validated using beatnote spectroscopy,
	a technique where the interferometric autocorrelation of the first order beating of the optical field is measured, using a Michelson interferometer and a fast detector. A Fourier transform of the resulting interferogram with sufficient resolution indicates the exact spectral regions contributing to the comb~\cite{hugi_mid-infrared_2012}, as well as highlighting those parts which are not locked. By demodulating the detected beatnote of a stabilised laser in quadrature, using as reference an RF tone at the same frequency $f_{rep}$, Burghoff and coworkers demonstrated that the individual phases could be recovered~\cite{burghoff_evaluating_2015}. These measurements were performed on a quantum cascade laser comb operating in the terahertz, and revealed that the device was lasing in a state that combined AM and FM. Because these devices are characterised by different operation (cryogenic vs room temperature) and physical parameters (gain recovery time order of 10s of ps for terahertz~\cite{green_gain_2009,bacon_gain_2016}, vs < 3 ps for mid-infrared~\cite{kuehn_ultrafast_2008,choi_gain_2008}), it was not expected that the behaviour observed in THz devices could be automatically assumed to apply in those operating in the mid-infrared. 
	

	\section{Methods}
	\label{sec:methods}
	
	For this study, we applied the coherent beatnote spectroscopy technique (also referred to as SWIFTS)\cite{burghoff_terahertz_2014} to a pair of near-identical mid-infrared devices, here labelled 21NU and 26HM. The latter are a pair of 6 mm dual-stack, high-reflectivity coated devices from the same process, lasing at 8.2 $\mu$m\cite{jouy_dual_2017-1} with up to a Watt of optical power. When operating as a comb, these devices are spectrally broad, lasing on hundreds of modes ($\sim$100 cm$^{-1}$). Stable comb operation is achieved using a waveguide featuring a small resonant coupling to the surface plasmon polariton propagating at the metal-semiconductor interface, such as to induce a controlled amount of negative GVD~\cite{bidaux_plasmon-enhanced_2017}. The resulting GVD is slightly negative overall, and equal to about -200 fs$^2$/mm in the laser spectral range (see Figure~\ref{Afig:GVD_21NU}). As such, it operates as a comb over several ranges of current, with strong beatnotes measured sub-kHz full width at half maximum. For these experiments, we operate the device at 291 K.
	
	\begin{figure}[!htb] 
		\centering
		\noindent\includegraphics[width=\linewidth]{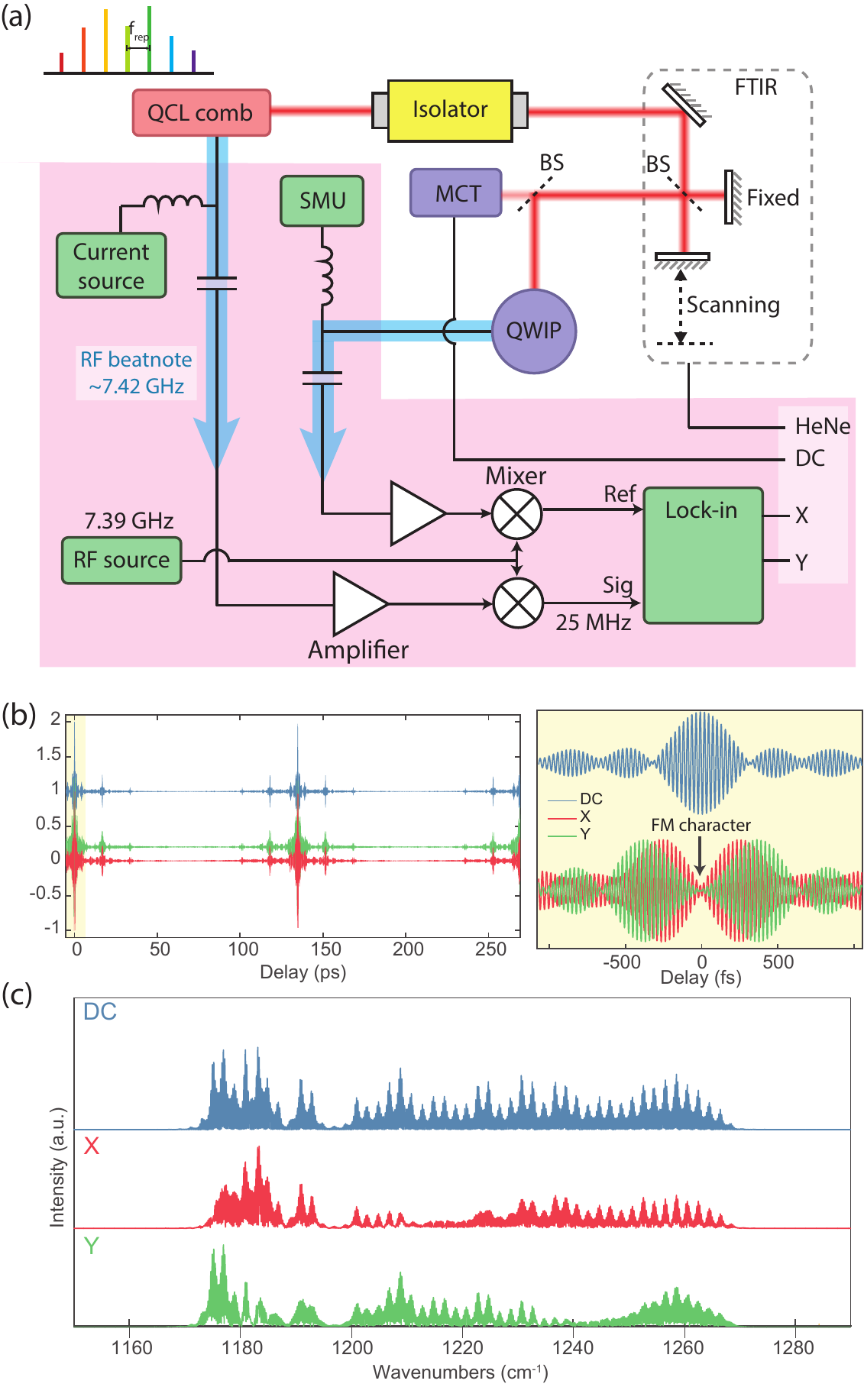}
		\caption{(a) SWIFTS\cite{burghoff_terahertz_2014} coherent beatnote spectroscopy setup, with detected signal referenced to QCL "clock". We measure with a maximum path difference corresponding to a resolution of 0.125 $cm^{-1}$, or about 3.75~GHz. The fast RF autocorrelation is measured on a Quantum Well Infrared Photodetector\cite{liu_qwip_1998} (QWIP), biased through a bias tee using a Source Measurement Unit (SMU, Keithley 2400). The reference and signal are amplified using a pair of low-noise, high-bandwidth RF amplifiers, and pre-downconverted to the lower frequency 25 MHz using Miteq mixers. The DC autocorrelation is collected on a separate, more sensitive Mercury Cadmium Telluride (MCT) detector.  (b) Example of measured DC and demodulated (SWIFTS) interferogram traces (left); zoom about centreburst (right), showing that, when the interferometer arms are balanced, the RF signal is at a minimum, characteristic of FM combs. (c) Computed spectra for the DC, X, and Y channels. }
		\label{fig:setup}
	\end{figure}
	
	A schematic drawing of the experimental setup is shown in Figure~\ref{fig:setup}. The light from the QCL is collimated by a lens (focal length 1.873 mm) before being sent through an optical isolator to a Fourier Transform Infrared Spectrometer fitted with a fast quantum well infrared detector (QWIP)~\cite{liu_qwip_1998}; this device has an electronic bandwidth well above 20~GHz, and as such the beatnote can be detected. On illumination, the beatings of the optical field will induce current modulations on the QWIP at the comb repetition rate $\omega_r$, and harmonics thereof. We write the field as:
	
	\begin{equation}
	E(t) = \sum_n A_n e^{i(\omega_0+n\omega_r t)}
	\end{equation}
	
	$n$ denotes the mode index, $\omega_0$ the carrier envelope offset frequency in radians per second, and $A_n=|A_n|e^{i\phi_n}$, with $\phi_n$ the average phase of mode $n$. The current fluctuations on the QWIP are then proportional to
	
	\begin{equation}
	\begin{split}
	\left \langle E^*(t)E(t+\tau) \right \rangle_{(\omega_r)} &= \sum_n A_n
	\Bigl [ A^*_{n-1} e^{i\omega_r t} + \Bigr . \\
	& \Bigl . A^*_{n+1} e^{-i\omega_r t}  \Bigr ] e^{i(\omega_0 + n\omega_r )\tau}
	\end{split}
	\end{equation}
	
	with $\tau$ the relative delay between the arms of the interferometer, which, for a rapid acquisition as in our case, becomes $\tau(t)$.  The angle brackets indicate an imagined narrowband filtering around the $\omega_r$, meaning we can focus on relevant terms; this is ultimately effected by the combined response of the mixers and the integration time on the lock-in.
	
	A reference for the beatnote, observed at around 7.4~GHz, is extracted from the device through the current feed of the QCL via a bias-tee~\cite{villares_dispersion_2016}. After amplification and downconversion by mixing it with a reference RF tone at $f_{LO}=f_r-25 MHz$, the $\sim$ 25MHz signal is injected into the reference channel of a fast lock-in amplifier from Zurich Instruments. The signal detected on the QWIP, after being equally downconverted using the same $f_{LO}$ tone, is then sent to the signal channel of the lock-in. Such a configuration, where the RF signal is referenced from the laser source itself, is advantageous as it mitigates drift, allowing us to measure the QCL free-running. Both quadratures of the detected signal, as retrieved by the lock-in, are recorded as a function of $\tau$. These quadratures are given by:
	
	\begin{equation}
	x(\tau) = \frac{1}{2}
	\sum_n A_n
	\left [ A^*_{n-1} +
	A^*_{n+1}  \right ] e^{i(\omega_0 + n\omega_r )\tau}
	\end{equation}
	
	\begin{equation}
	y(\tau) = \frac{1}{2 i}
	\sum_n A_n
	\left [A^*_{n-1} - A^*_{n+1} \right ] e^{i(\omega_0 + n\omega_r )\tau}
	\end{equation}
	
	As shown in Figure~\ref{fig:setup}~(b), these two interferograms are recorded in parallel with the normal DC autocorrelation signal, which is instead measured on a more sensitive, low bandwidth detector.
	
	As clear from Equation~\ref{eq:phases}, the phasors $|A_n||A_{n-1}|e^{i(\omega_r t+\phi_n-\phi_{n-1})} $ are retrieved by a simple Fourier transform of both interferograms, allowing the computation of the inter-modal phase differences:
	
	\begin{equation}
	X(\omega)+ i Y(\omega) =  \sum_n{|A_n| |A_{n-1}| e^{i(\phi_n-\phi_{n-1})} \delta \left [\omega-(\omega_0 + n\omega_r) \right ]}
	\label{eq:phases}
	\end{equation}
	
	$\angle{(X+iY)}=\phi_n - \phi_{n-1}$ can be understood as the group delay multiplied by the mode spacing $\omega_r$. The phases themselves are found by a cumulative summation, assuming one arbitrarily selected mode, here written $n=0$, to have a phase of zero\cite{debeau_simple_1998}:
	
	\begin{equation}
	\phi_n = \begin{cases}
		\sum_{n'=0}^n -(\phi_{n'}-\phi_{n'-1}) & n<0\\
		0 & n=0\\
		\sum_{'n=0}^n \phi_{n'}-\phi_{n'-1} & n>0\\
		\end{cases}
	\end{equation}

	The positive amplitudes are found by simply taking $\sqrt{|An|^2}$ from the normal power spectrum, which is acquired in parallel. We address the data treatment in more detail in Appendix~\ref{Asec:data_treatment}.
	
	\section{Results}
	\label{sec:results}
	
\begin{figure}[!htb]
		\centering
		\includegraphics[width=\linewidth]{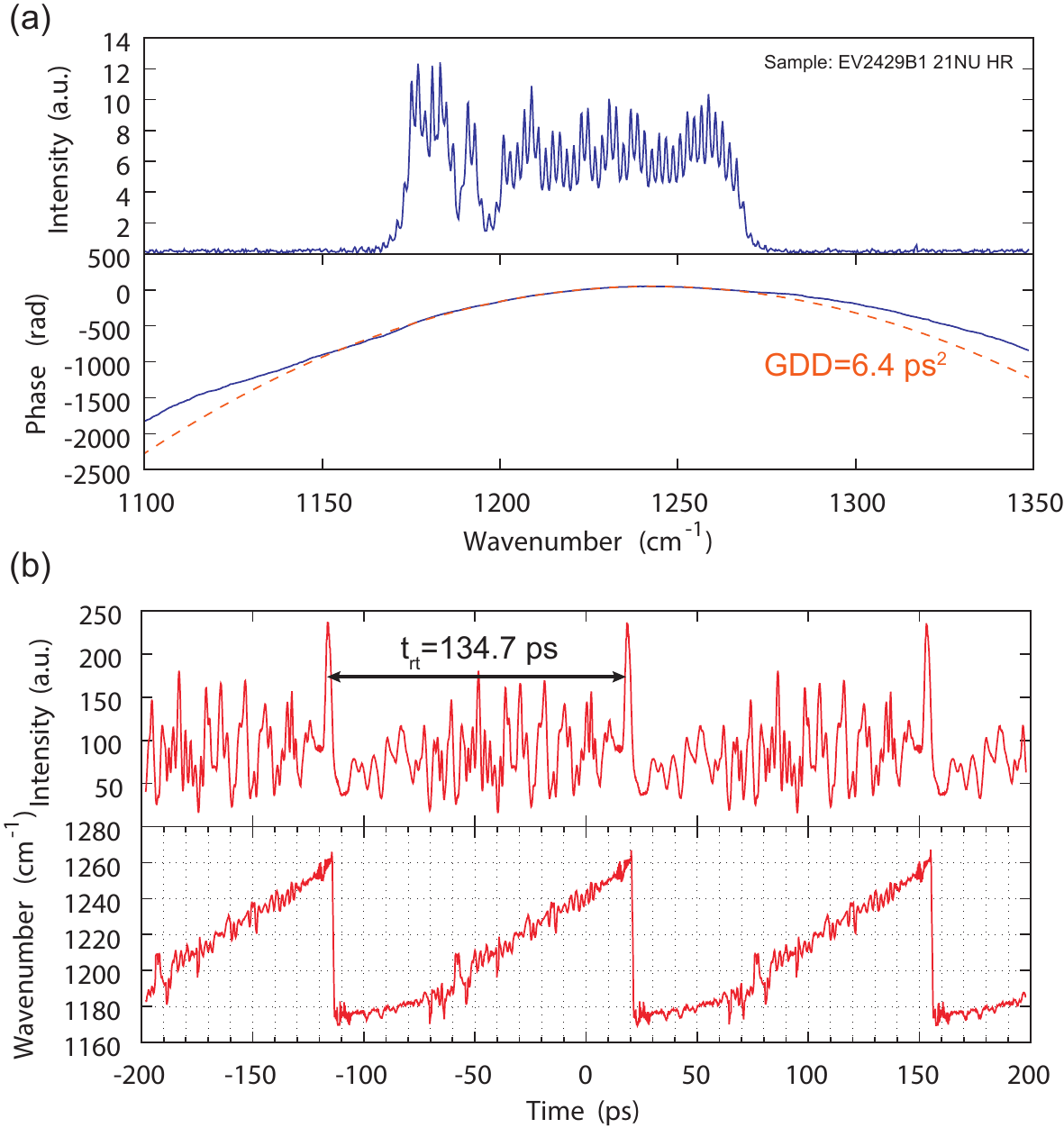}
		\caption{(a) Amplitude spectrum and computed phase for one measurement set. The orange dashed line indicates the quadratic fit to the phase, yielding an estimated GDD of 6.4~ps$^2$. (b) instantaneous intensity and frequency of the field, simulated for the measurement state in (a). Both have been integrated to 1 ps to reflect the response times of the active region.}
		\label{fig:time_domain}
	\end{figure}

	We focus here on the results for one device, the 21NU; data for the 26HM sample is available in the Appendix~\ref{Asec:Amore_data}. The modal amplitudes and unwrapped phases are shown in Figure~\ref{fig:time_domain}~(a), revealing a parabolic phase profile corresponding to a GDD of $-6.4$ps$^2$. Knowledge of the amplitude and phases enabled us to compute the temporal dependence of intensity as well as the instantaneous frequency of the device during one period of the comb, which are presented in Figure~\ref{fig:time_domain}~(b). It is observed that the laser operates with a non-negligible amplitude modulation with a very short characteristic time, contrary to prediction. However, the most striking aspect of the result is the simple linear chirping behaviour, where the emission frequency is monotonically increasing from the lower frequencies to the higher ones. 

	Intuitively, one can understand the correspondence between the parabolic phase profile and the linear chirp by considering the group delay. In this case, the field sweeps a bandwidth of $f_{BW}$ during one single period of the comb $f_{rep}^{-1}$, yielding a group delay of $GD(\omega)=(\omega-\omega_0)/f_{BW}f_{rep}+\Psi$, with $\Psi$ representing an arbitrary delay offset, with imparts no distortion on the signal shape. The expression is linear in frequency; the phase, which is the integral of this with respect to the circular frequency, will then naturally yield a parabolic phase profile. This parabolic phase profile is then characterised by a group delay dispersion given by:
	\begin{equation}
	\textrm{GDD} = \frac{1}{2 \pi f_{rep} f_{BW}}.
	\label{eq:gdd}
	\end{equation}  
	This simple argument yields a predicted GDD of $-7.1$ps$^2$, close to the value measured of $-6.4$ps$^2$.
	
	\begin{figure}[!htb]
		\centering
		\includegraphics[width=\linewidth]{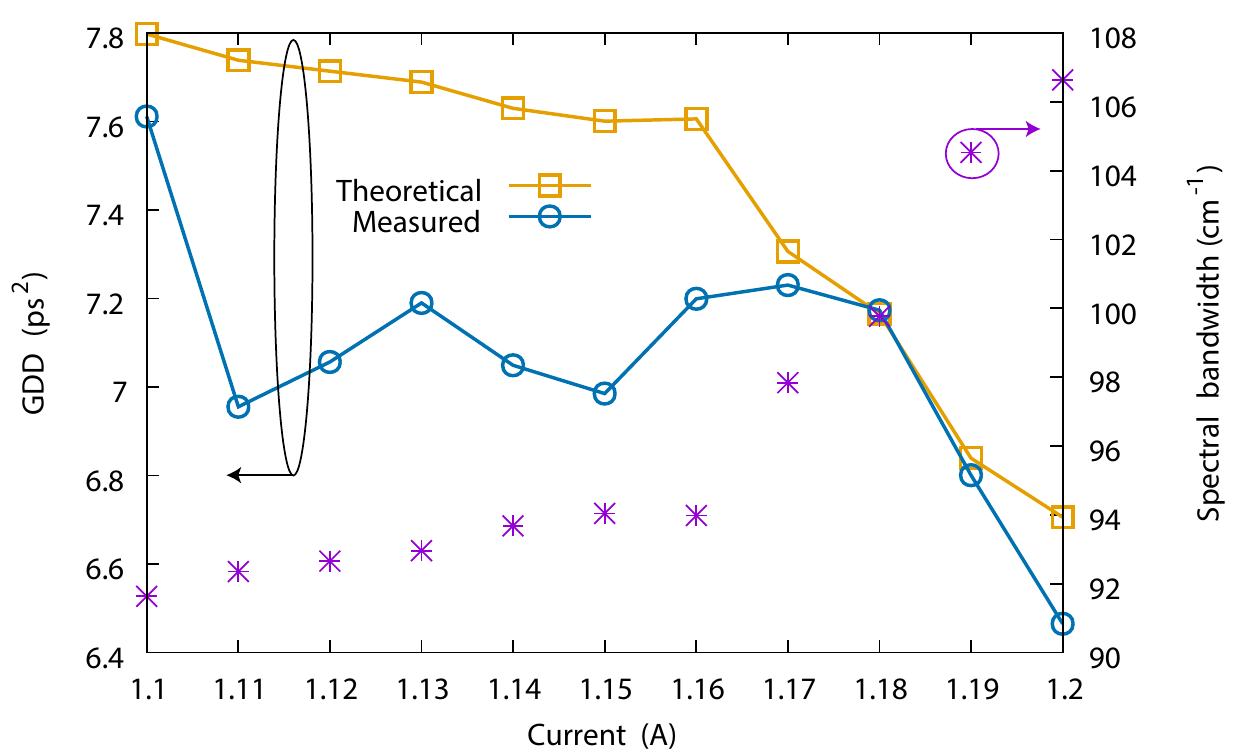}
		\caption{GDD of the laser output field, as predicted by Equation~\ref{eq:gdd} from the spectral bandwidth and round trip frequency (orange), and estimated by parabolic fit to unwrapped phase profile (blue). The dashed purple line indicates the estimated spectral bandwidth at each value of current.}
		\label{fig:coarse_tuning}
	\end{figure}

	We assessed the current dependence of the phases by firstly tuning the QCL into a comb state, and by then slowly ramping the current, acquiring the spectra in 10 mA increments. Over this range, the field is seen to retain its linear group delay characteristic, with the spectrum for the most part broadening towards the blue, and filling out. This suggests that the chirped comb state is by no means fragile, and that the phase in this case tunes smoothly as a function of the current.
	
	To see how well the linear chirped state describes our field, we compare the GDD predicted in Equation~\ref{eq:gdd}, taking estimates of the spectral bandwidth and repetition rate, to parabolic fits on the measured phase. As shown in Figure~\ref{fig:coarse_tuning}, we do indeed see a decrease in the GDD with increasing current (bandwidth), and the agreement at around 1.18 A is very good. Such behaviour is particularly interesting from the perspective of pulse generation in that, when the spectrum is broader, the field dispersion is in a sense easier to compensate. We attribute the discrepancy between the two curves mainly on two things. Firstly, in taking the spectral bandwidth estimate at -15 dB from the peak, we are in general under-evaluating the bandwidth in exchange for a more reliable estimate. Secondly, there is a clear presence of higher order dispersion, especially towards the extrema of the spectrum (see Appendix~\ref{Asec:Amore_data}). This effect is even more apparent for the two spectral lobes in the 26HM sample. We attribute this partly to the cold-cavity dispersion, which is observed to cross zero at the lower frequency side, about where the spectrum drops in intensity, and which again seems to have a turning point about where the spectrum is growing in the higher frequency direction. Somewhat surprisingly, this seems to have the effect of increasing the GDD in these regions.
	
	To assess the reproducibility of the chirped state, two complementary measurements were made. In the first, the device was switched on, the current ramped up in 2.5 mA steps, and the beatnote frequency and power recorded at each step. Such a measurement was repeated 37 times, and, in the forwards direction, the frequency and power of the beatnote was identical in all cases within experimental uncertainty (see Figure~\ref{Afig:bn_trace}). This result strongly suggests that the final comb states for this laser are deterministic across all values of current, when the current is increased in a controlled manner; however, more work in the direction of time-resolved studies is required before we can understand whether the excitation pathway itself is also deterministic.
	
\begin{figure}[!htb]
		\centering
		\includegraphics[width=\linewidth]{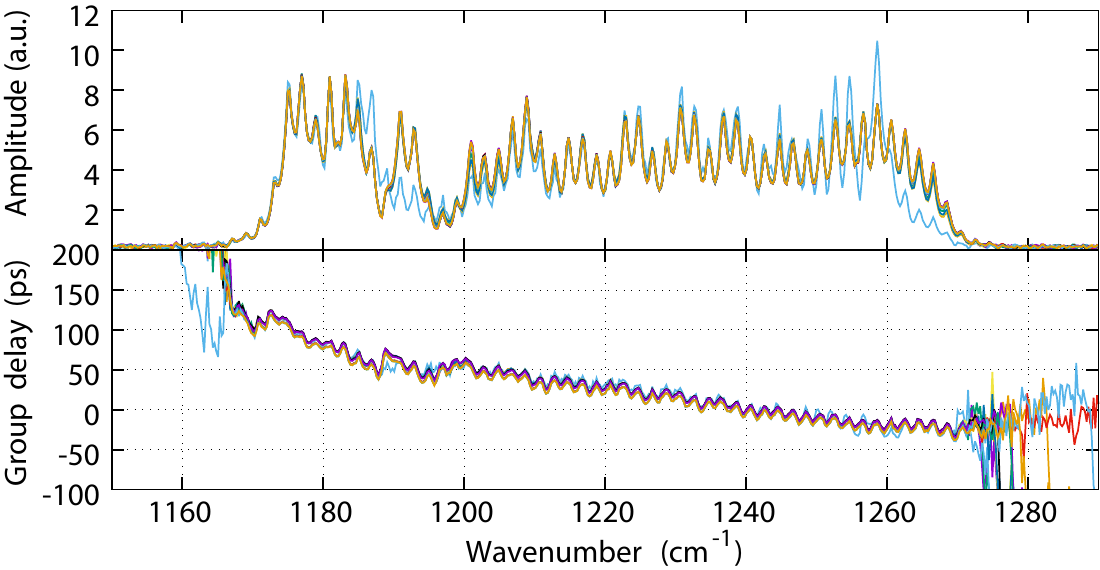}
		\caption{12 amplitude and group delay traces, with the power having been cycled prior to measuring each. Vertical offsets in the group delay traces are unimportant, as they only represent relative time delays.}
		\label{fig:power_cycling}
	\end{figure}

	In the second measurement, spectral amplitude and phase measurements were made repeatedly as before, with the power being cycled in-between. The amplitude spectra and corresponding group delays presented in Figure~\ref{fig:power_cycling} are for 12 cycles of power, and the results show very good agreement amongst themselves, confirming the comb state was reproduced. Notably in one case, the spectral characteristic was much different, suggesting the measurement was affected either by optical feedback, or perhaps a stochastic effect if we were measuring close to a multistability. Remarkably, in spite of the different amplitude distribution, the group delay structure is preserved.
	
	 Having obtained these results, it is reasonable to ask why such a quadratic phase state should be selected. The approach we take here is to note that such a state should also be a favourable solution of the Maxwell-Bloch coupled mode equations which govern the dynamics of our device. To this end, we write~ \cite{khurgin_coherent_2014,villares_quantum_2015}
	
	\begin{multline}
	\frac{dA_n}{dt} = (G_n -1)A_n -iD_n A_n \\- G_n\sum_{k,l=-N/2}^{N/2} A_m A_k A_l^* B_{kl}C_{kl} \kappa_{klmn} 
	\label{eq:modedyn} 
	\end{multline}
	
	for the mode $n$ of amplitude $A_n$ experiencing a gain $G_n$ and a phase shift $D_n$ due to the cavity dispersion. $\kappa$ is a coefficient describing the axial overlap of modes $k,l,m,n$, and $m=n+k-l$ for convenience. In Equation~\ref{eq:modedyn} the modes interact via the four wave mixing term where the $B_{kl}$ and $C_{kl}$ represent the phase and population pulsation and $\kappa_{klmn}$ the modal overlap, all as defined in Appendix~\ref{Asec:merit}. In the comb regime, the mode amplitudes are constant and their time derivative zero (i.e. $\dot{A}_n = 0$). 
	
	We note now that, in the absence of significant cavity GDD, neither the first nor the second term of the right hand side of \ref{eq:modedyn} will introduce a strong GDD. Being negative, the four wave mixing term favours solutions that possess chirp, which then minimises its absolute value, but will not introduce a strong GDD.
	
\begin{figure}[!htb]
		\centering
		\includegraphics[width=\linewidth]{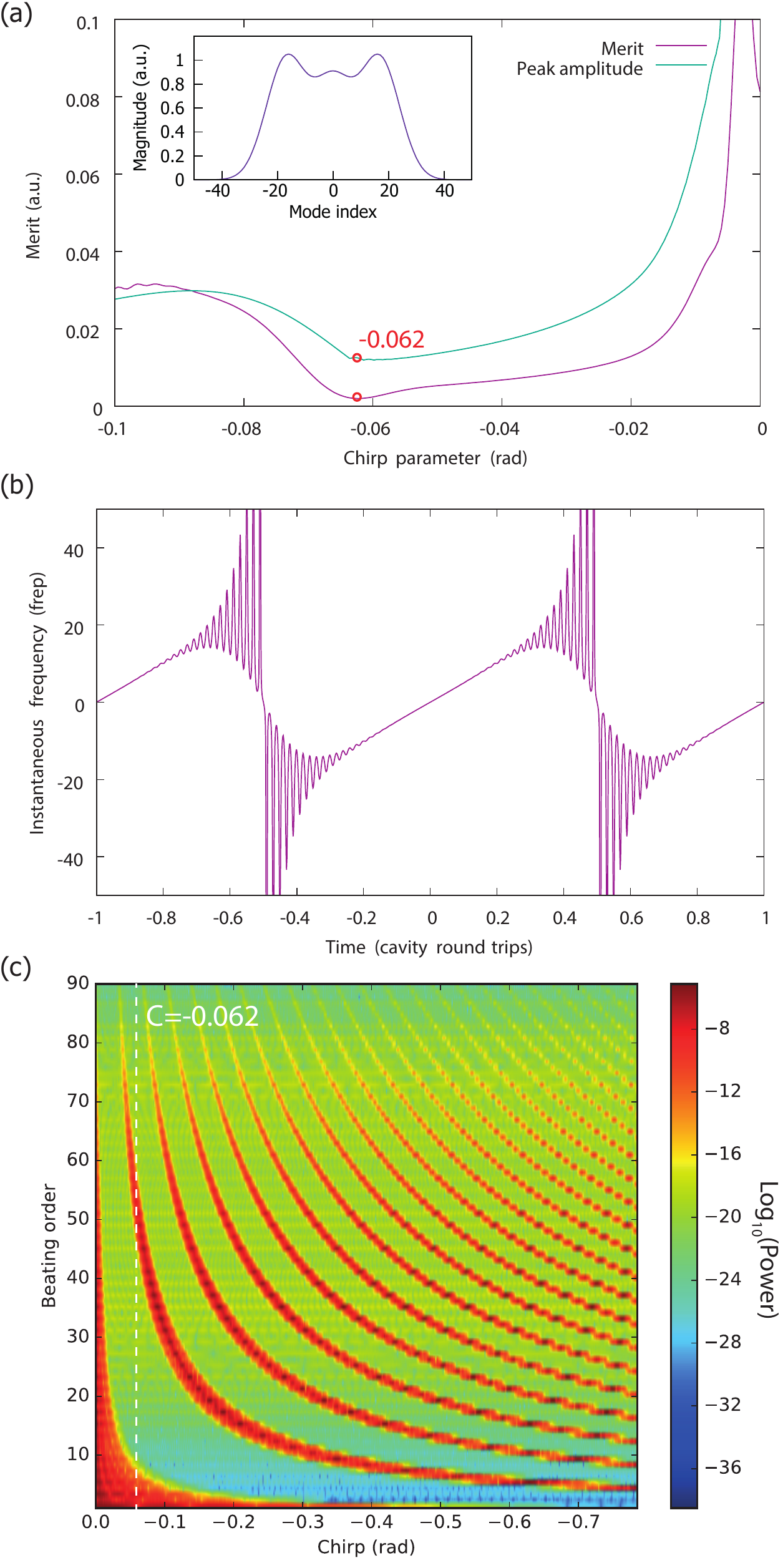}
		\caption{(a) Merit as a function of the chirp parameter. Parameters: $\tau_{21}=0.5$ ps, $\tau_{22}=0.1$ ps, $\tau_{rt}=133$ ps. The red line at about -0.062 indicates the first minimum, which coincides with the first minimum of amplitude modulation. Inset: spectral amplitude as a function of mode index, chosen to be a sum of 3 Gaussians. $\sigma=10$ modes, $A=[1,0.8,1]$, offset 17, $g_0=1.12$.  (b) Frequency as a function of time for the state with the best merit function, showing a linear chirp. (c) Beats present in the intracavity field for a given chirp (see Appendix~\ref{Asec:dissipation}). The dashed line corresponds to our optimal chirp, and shows the onset of strong high frequency beat tones.}
		\label{fig:theory}
	\end{figure}

	To support these qualitative considerations, we tried a variational approach to solving equation~\ref{eq:modedyn}, treating the absolute value of the r.h.s. as the merit function for a given amplitude, parametrised by its chirp. As shown in the inset of Figure~\ref{fig:theory}~(a) we choose a generic form for the amplitude which roughly describes the one we observe experimentally. We then assume that the phases are parametrised by a chirp parameter $c$ that yields a phase $\exp( i c n^2)$, where $n$ is the mode index, relative to the primary mode.  Figure~\ref{fig:theory}~(a) shows the merit as function of $c$, which exhibits a minimum for the value $c = -0.062$ indicated by the red line. As illustrated in Figure~\ref{fig:theory}~(b), this value of chirp corresponds to an instantaneous frequency which increases for the most part linearly in time.
	
	This result was observed not to depend critically on the precise shape of the amplitude function, nor on $\tau_{22}/\tau_{rt}$. We note however that this argument does not exclude the existence of more highly chirped linear states, which can be found at chirps of, for example, ${c=2\pi (1\pm1/N)/m}$, where $m$ is some integer. Indeed, calculating the merit function beyond -0.1~rad yields a series of minima coinciding with these points (see Appendix~\ref{Asec:dissipation}). It remains to be seen why this chirp should be chosen among others, but we postulate that a loss acting on higher order beatings, which modulate the gain through photon driven transport\cite{piccardo_time-dependent_2018}, could render lasing in such a configuration unfavourable; these high frequency beatings start to appear strongly around this first minimum (see Figure \ref{fig:theory}(c)), where the pulse is becoming sufficiently stretched that its short wavelengths start to interfere with the long wavelengths of the previous pulse.
	
	\section{Conclusion}
	\label{sec:conclusion}
	By measuring the inter-modal phases, we have shown that mid-infrared QCLs can exhibit a temporal profile which is approximately described by a linear chirp. We have found this state to be deterministic and stable, with all parameters tuning smoothly with the current. This is encouraging from several perspectives. Firstly, the field GDD can be compensated to the first order in a relatively simple fashion (using for instance a dispersive fibre or fixed grating pair), yielding short, powerful pulses; such a pulsed field could then be used in non-linear spectroscopy, for example. Secondly, we can now work to deepen our understanding of why such a state should result. From a state engineering perspective, one could eventually hope to design or steer the comb to a state which is broader, more robust, or perhaps with other more desirable temporal characteristics.
	
	We used a simple model based on the Maxwell-Bloch coupled mode equation to explore the feasibility of quadratic phase states. We've found that for specific chirp values, the absolute value of the time derivative is minimised, and indeed the optimal chirp by this measure coincides with the linear phase profiles for which AM is minimised. This simple study does not rule out the possibility for higher order chirp states, at which the merit function is also minimised, or indeed other more exotic configurations, such as those classified pseudo-random FM. Nonetheless, we note that similar chirping behaviour has also been observed in a range of self-pumped FWM combs, including single-section Fabry-P\'erot Q-dash~\cite{rosales_high_2012}, and Q-well lasers, both measured \cite{sato_100_2001} and simulated \cite{dong_traveling_2017}, suggesting that this may be a more general behaviour of such systems. The precise role of system parameters, including dispersion, leading to this simple chirp, we hope to clarify at a later date.
	
	\section*{Acknowledgements}
	We would like to express our gratitude to Dmitry Kazakov, Zhixin Wang, and Dr. Elena Mavrona for their careful reading of the manuscript and helpful comments. This work was financially supported by the Swiss National Science Foundation (SNF200020-165639).

	
	\appendix
	\section{Acquisition and data treatment}
	\label{Asec:data_treatment}
	A 7.3 GHz RF tone (Rohde and Schwarz SMF 100A) is used as the local oscillator (LO) to downmix the amplified beatnote derived from the QCL, and the signal from the QWIP; these are then further downmixed and integrated on a fast (50~MS/s) lock-in amplifier (Zurich Instruments HF2LI), with an integration time corresponding to $\tau\approx0.5\lambda_0/v$ ($v$ the mirror velocity and $\lambda_0$ the central wavelength). A stabilised helium neon (HeNe) laser is also aligned to the interferometer (Brucker IFS 66/S), allowing the mirror displacement to be measured. This HeNe signal, the DC autocorrelation taken from the MCT, and lock-in signals are acquired on a 4-channel fast oscilloscope (LeCroy HDO6104). We generally oversample as far as the memory allows us, and then subsequently digitally filter the traces for a processing gain in SNR of a factor $\sqrt{N}$ for an oversampling factor $N$.
	
	Care is taken to keep the coaxial cable lengths for the X and Y channels the same, as well as to use the identical filters, so as not to induce any artificial phase shifts between the components. Synchronisation between the DC autocorrelation and RF autocorrelation traces is achieved by comparing the positions of the centreburst (zero path difference) in the DC autocorrelation and the smallest peak in $|x(t)+i y(t)|^2$. As there is an AM component to the field oscillating at the repetition rate, as made clear by the existence of a beatnote when light is shined directly onto the detector, such a peak must exist.
	
	Zero crossings of the HeNe signal are identified, and linear interpolation is used to improve the precision in their estimate. The 3 single-sided interferograms are then resampled accordingly using a cubic interpolator. We thus move from units of the scope's internal LO to the relative path delay. Being considered narrowband signals (fractional bandwidth < $100 cm^{-1}/1200 cm^{-1}\approx8.4\%$), phase correction steps are deemed unnecessary.
	
	The interferograms are then apodised using the Mertz window~\cite{mertz_auxiliary_1967}, zero-padded, left circular shifted to the centreburst, and Fourier transformed, as is conventionally done for single-sided interferograms. A first estimate of the comb repetition rate is then found either by adding the lock-in demodulation frequency to the LO frequency used for the downconversion, which is usually selected to be below the fundamental beat tone of the QCL, or by measuring the delay in the interferogram from the centreburst to the first satellite. The estimate is then improved iteratively on the DC autocorrelation spectrum, which generally has a higher SNR, by choosing the values of $\hat{\omega_0}$ and $\hat{\omega_r}$ which maximise $\sum_n{I(\hat{\omega}_0+n\hat{\omega_r})}$.
	
\begin{figure}[!htb]
		\centering
		\includegraphics[width=\linewidth]{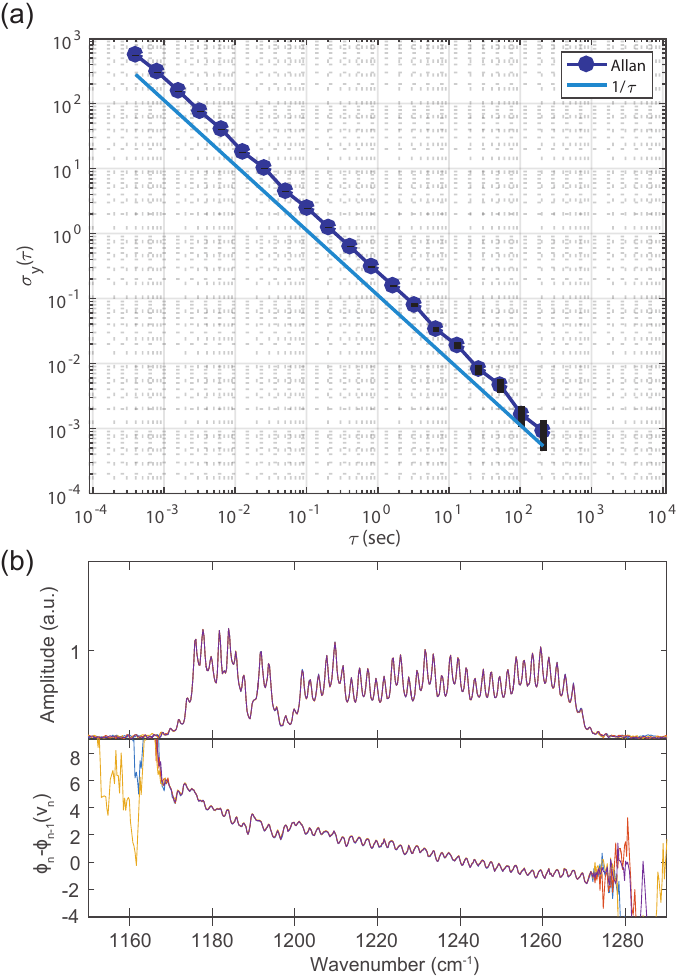}
		\caption{(a) Allan deviation measured with the mirror at a fixed position, for a maximum delay of 100 s. (b) Measurements of the phase differences taken at a single operating point, with no power cycling between, and treated identially.}
		\label{Afig:stability}
	\end{figure}

	For stability, we made use of a low-noise current driver (Wavelength Electronics QCL2000) and thermoelectric cooler (Wavelength Electronics PTC 10 K-CH, with a standard Peltier element), in combination with an RF self-referencing scheme to mitigate drift. To verify the stability, we measured the deviation of the signal from the lock-in at a fixed path difference at rapid intervals over the course of more than 20 minutes. As shown from the Allan deviation\cite{allan_statistics_1966} in Figure~\ref{Afig:stability}~(a), there is no obvious drift in the signal over 100 seconds. The stability is confirmed in~(b), where superimposed are 13 amplitude and phase difference traces, for measurements taken sequentially at the same operating point with no power cycling in-between. As can be seen, they are very consistent and one can legitimately average.
	
	\section{Samples and further analysis}
	\label{Asec:Amore_data}
	\begin{figure}[!htb]
		\centering
		\includegraphics[width=\linewidth]{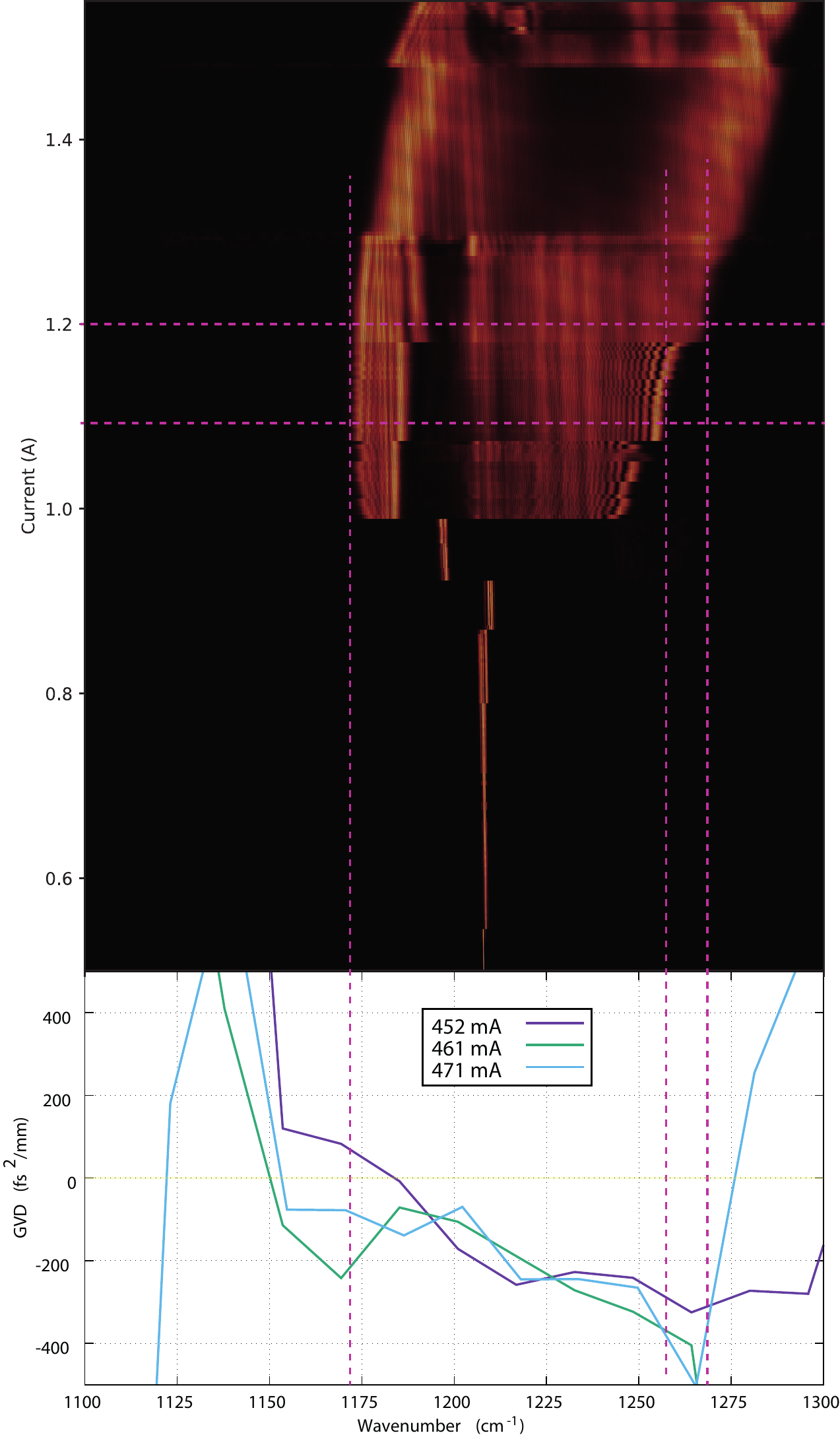}
		\caption{GVD of device measured subthreshold at 298 K using the Hakki-Pauli method\cite{hofstetter_measurement_1999} at 3 different current setpoints. The green lines indicate the approximate spectral width where the phases were measured. The pink dashed lines indicate the range of currents (1.10-1.20 A) over which the majority of measurements were made for the 21NU device, and the spectral bandwidth they cover.}
		\label{Afig:GVD_21NU}
	\end{figure}
	
	\begin{figure}[!htb]
		\centering
		\includegraphics[width=\linewidth]{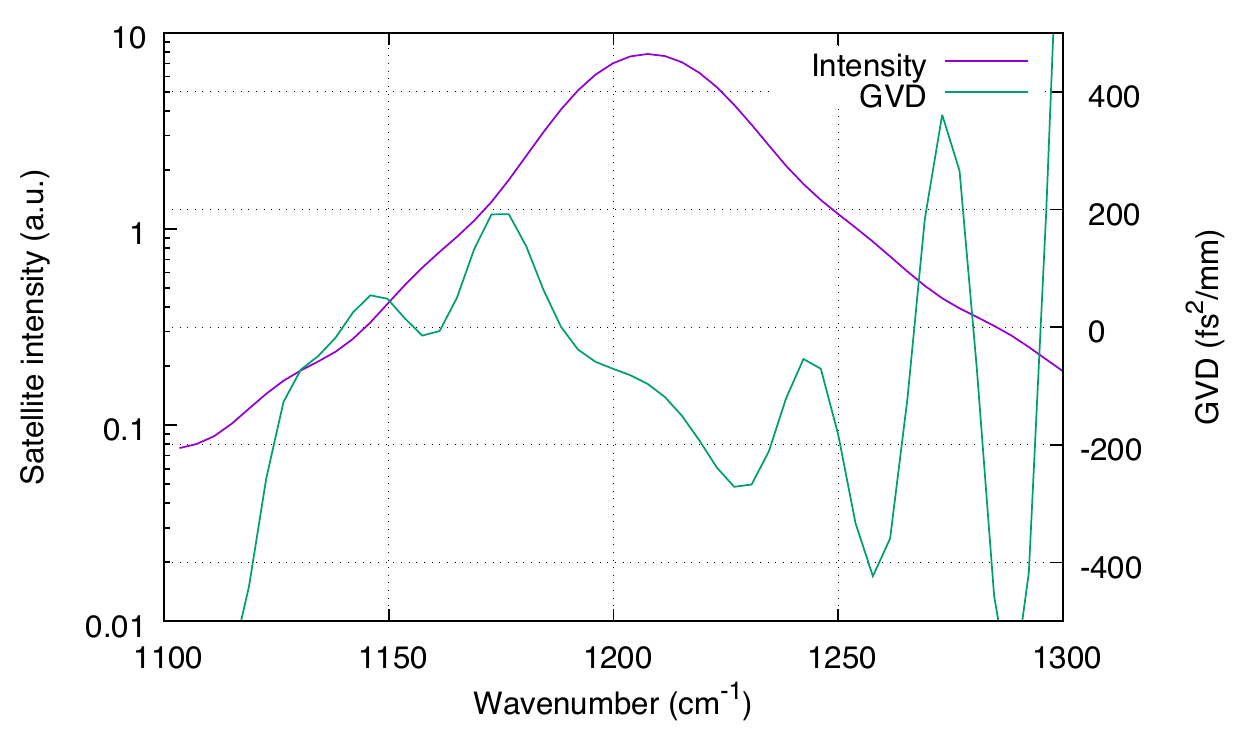}
		\caption{GVD for the 26 HM device, taken just below threshold. The purple line indicates the intensity of the spectrum of the first satellite, giving an indication of the range over which the measurement can be trusted. The blue curve gives the GVD.}
		\label{Afig:GVD_26HM}
	\end{figure}
	
	Two similar devices were characterised for this paper, of the same active region EV2429, and process B: 21NU, and 26HM. Both devices have a high-reflectivity coating on the back facet (300 nm Al2O3, 150 nm Au). The GVD for 21NU is shown in Figure~\ref{Afig:GVD_21NU}. It varies from around -200 to -400 fs$^2$/mm, and appears to cross zero at around 1175 cm$^{-1}$. At 1275 cm$^{-1}$ there seems to be another turning point. The values for the 26HM Figure~\ref{Afig:GVD_26HM} show similar features and values.
	
\begin{figure}[!htb]
		\centering
		\includegraphics[width=\linewidth]{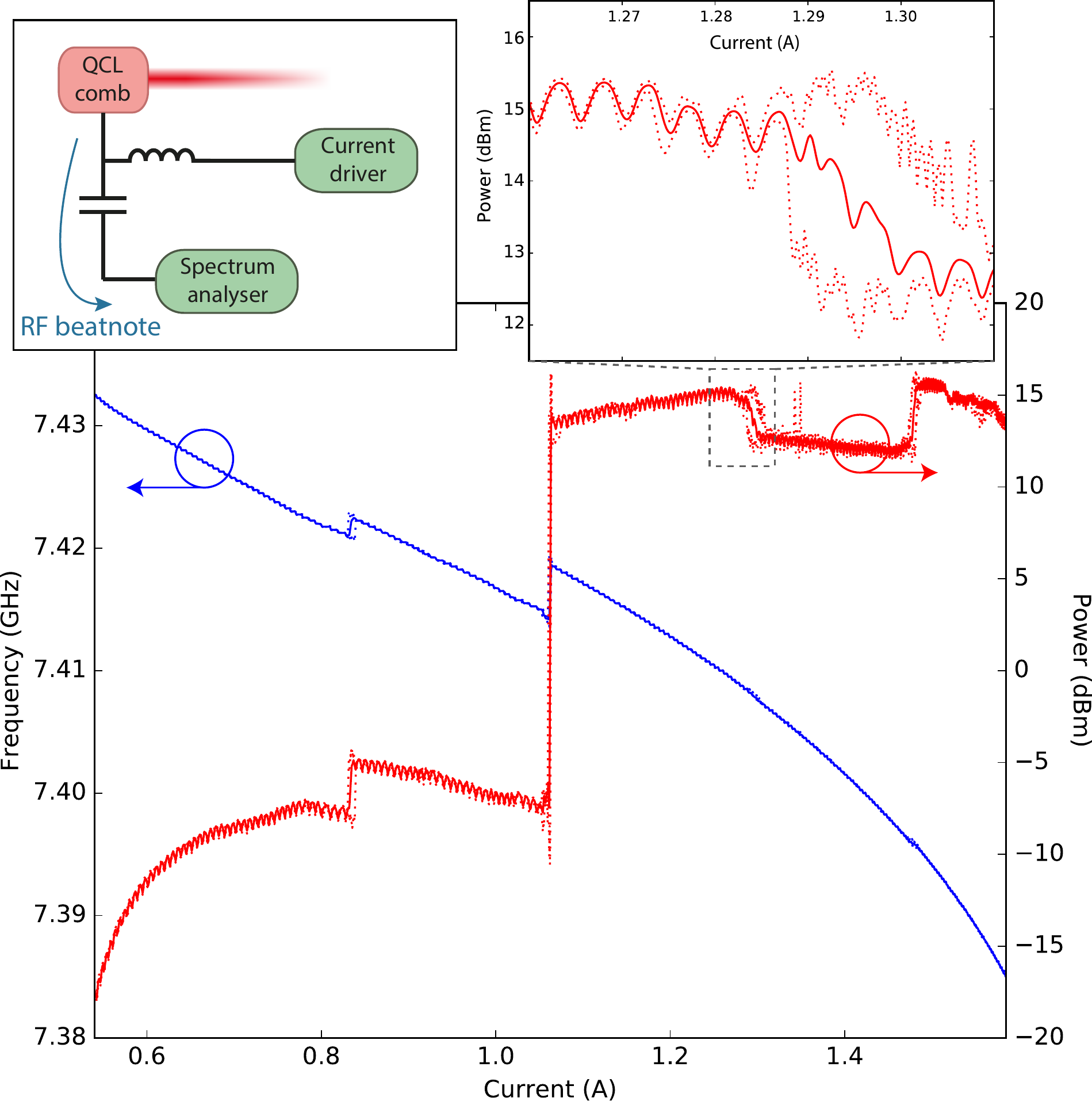}
		\caption{Average (solid) and minima/maxima (dotted) lines show the frequency (blue) and power (red) in the beatnote, measured for 37 current ramps. Inset left: setup used to drive the laser and measure the beatnote. (Spectrum analyser: Rohde and Schwarz FSU 50). Inset right: power traces zoomed at 1.28 A, showing both consistency (small range, left) and a more noisy region (larger range, right).}
		\label{Afig:bn_trace}
	\end{figure}

	An additional measurement noted in Section~\ref{sec:results} is shown in~\ref{Afig:bn_trace}, where the current is ramped in small steps of 2.5 mA and the beatnote frequency and power, as measured at the bias tee connected to the device under test, are recorded. As can be seen, the result is identical within experimental error on for each of the 37 traces; small inconsistencies between the traces can largely be attributed to differences between the current setting and acquisition time.
	
	In the reverse direction, there is generally good overlap, but the laser instead converges to one of several discrete levels of power and frequency. This suggests the existence of several multistabilities, and generally agrees with the observation of hysteresis as noted in \cite{mansuripur_qcl_nodate} 
	
\begin{figure}[!htb]
		\centering
		\includegraphics[width=\linewidth]{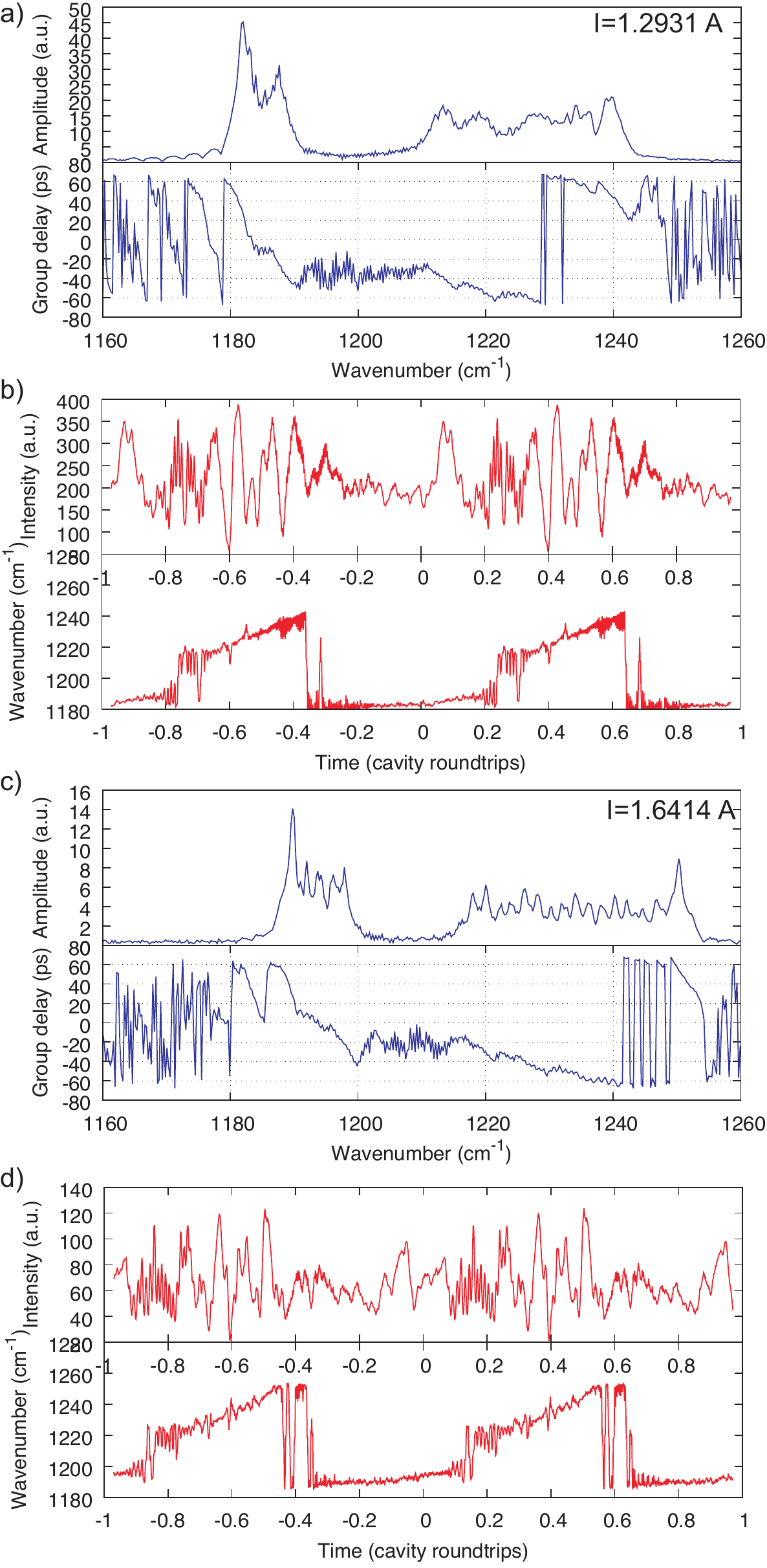}
		\caption{(a), (c) Modal amplitudes and phase differences, scaled to the comb repetition rate to give the group delay, for 26HM, measured at 1.2931 A and 1.6414 A, respectively. (c) and (d) Extracted intensity and instantaneous frequency. Note that, for these traces there was no averaging, and so the SNR is generally poorer, as most apparent near the centre of the spectrum. }
		\label{Afig:field_26HM}
	\end{figure}

	The second device, 26 HM, also demonstrates this chirping behaviour, with the difference in group delay between the left and the right lobes also appearing. Measurements of the modal amplitudes and corresponding group delay are shown in Figure~\ref{Afig:field_26HM}~(a) for 1.2931 A and (c) for 1.6414 A. As before, the time dependent intensity and instantaneous frequency are also plotted in (b) and (d). This further shows that the chirped state persists, or at least exists at multiple points, even over a larger current range than that measured for the first device.
	
\begin{figure}[!htb]
		\centering
		\includegraphics[width=\linewidth]{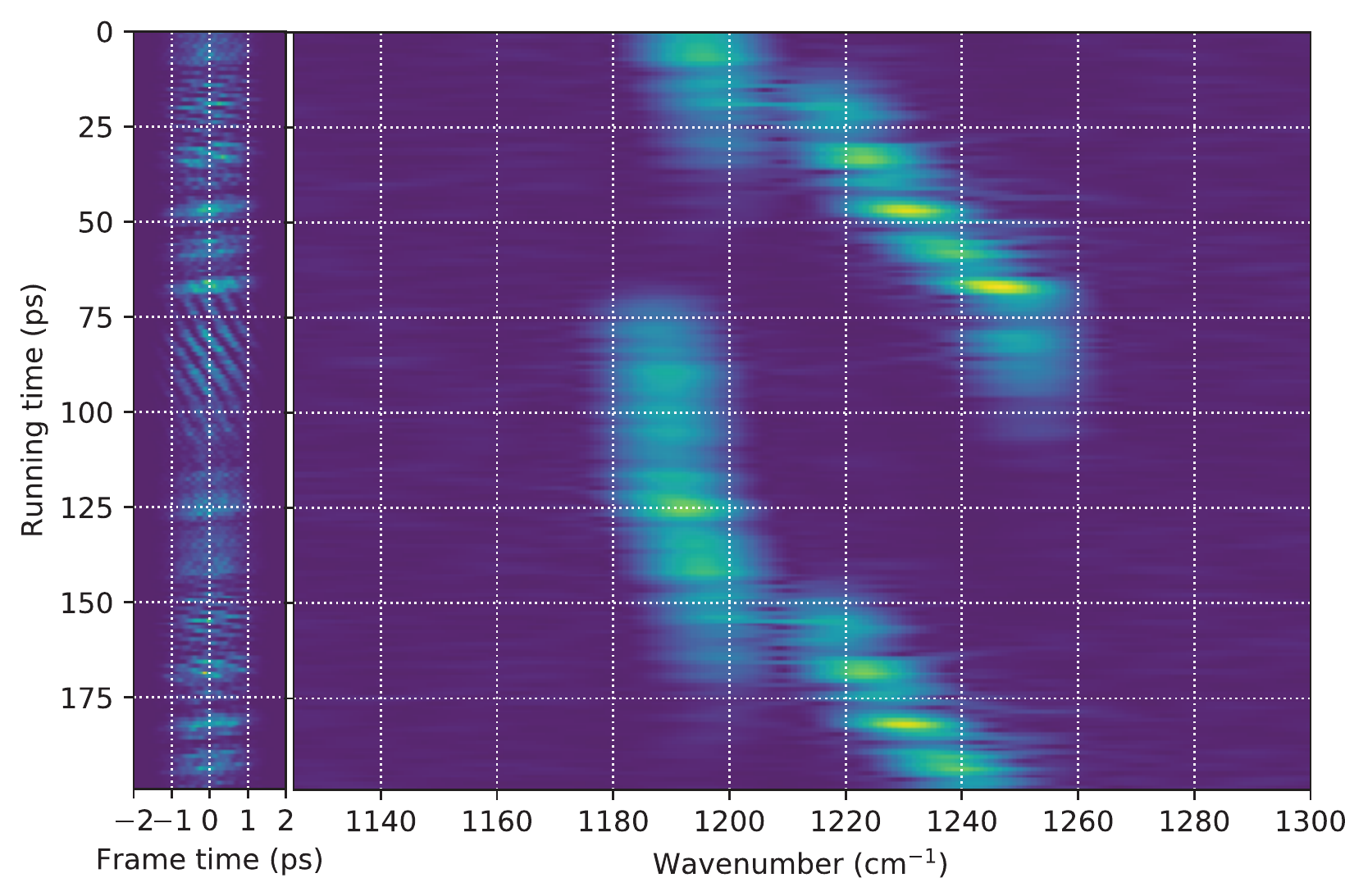}
		\caption{Spectrogram computed for Figure~\ref{Afig:field_26HM}~(d). Left: time slices (~5 ps frame time). Right: Corresponding Fourier spectrum.}
		\label{Afig:spectrogram_26HM}
	\end{figure}

	Note that the plotted instantaneous frequency in places shows rapid swings between two extrema. This is simply an artefact of the 1D representation, and corresponds to multiple separated spectral components with similar group delay. The spectrogram in Figure~\ref{Afig:spectrogram_26HM}, shows exactly this behaviour, with the two distinct GDDs for the two lobes plainly visible, and the overlapping components at around 25 ps.
	
\begin{figure}[!htb]
		\centering
		\includegraphics[width=\linewidth]{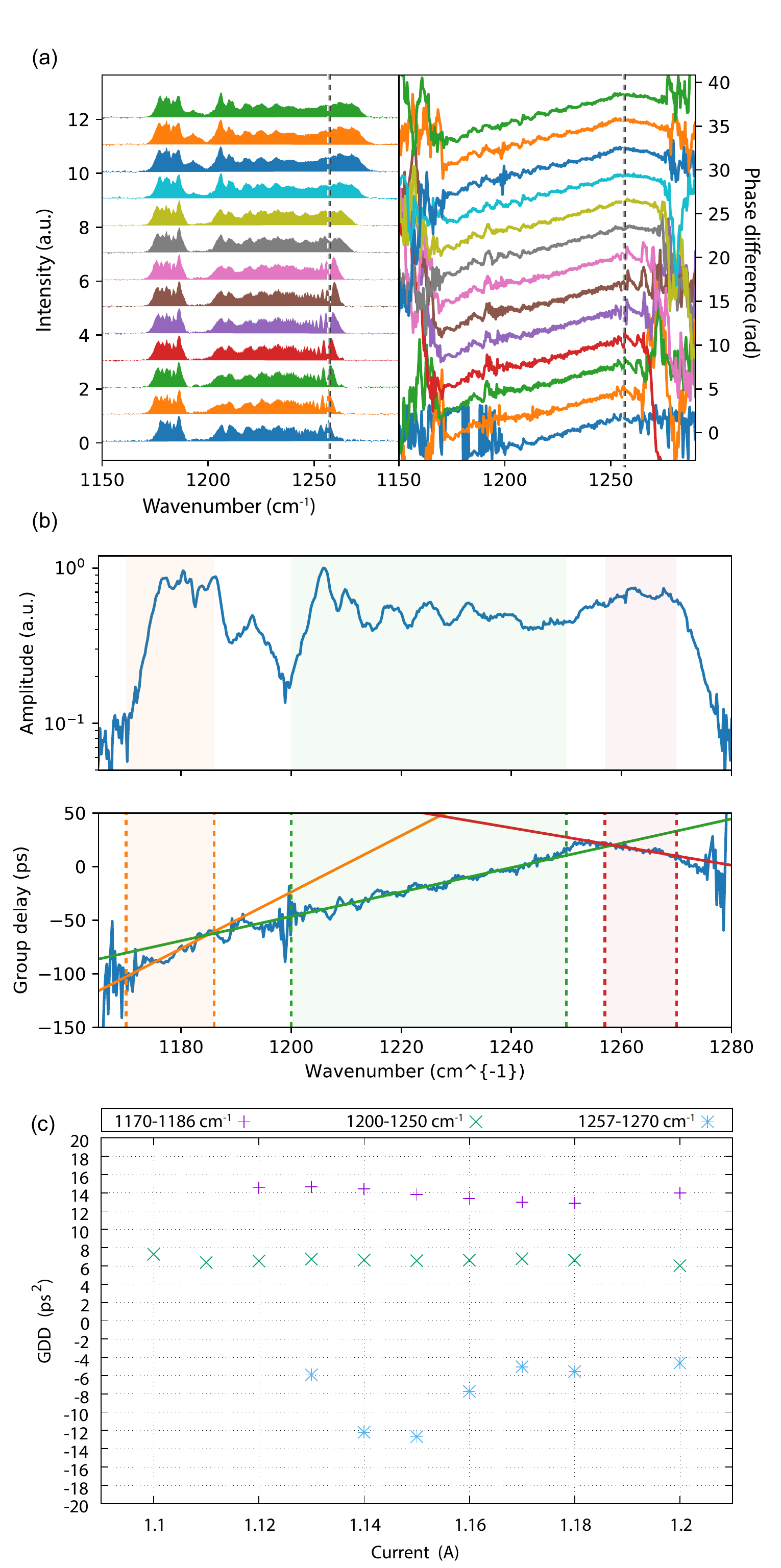}
		\caption{(a) Amplitude spectrum (left) and phase differences (right) measured as a function of current for device 21NU. The current increases in 10 mA steps, going from the bottom up. The blue dashed line indicates where the GVD changes from negative from positive. (b) Linear fit of group delay at the lobes 1170-1186 cm$^{-1}$ (orange), 1200-1250 cm$^{-1}$ (green), and 1257-1270 cm$^{-1}$ to estimate the GDD of the 3 distinct regions, plotted for 1.20 A. (c) Corresponding GDD estimates for all current values.}
		\label{Afig:coarse_tuning}
	\end{figure}

	Such a temporal response shows it something of an oversimplification to describe the field as a plain linear chirp. Plotted in Figure~\ref{Afig:coarse_tuning}~(a) are the intermodal phase differences, from which the overall GDD was estimated for the 21NU device. Looking more closely, in particular at the highest current values, one can see that there is in fact higher order dispersion present in the field. This seems to manifest itself in three distinct regions: the lobe to the left of 1200 $cm^{-1}$; the broader region to the right of 1200 $cm^{-1}$; and those new frequency components which grow from the main comb towards the blue. In (b), we perform 3 crude linear fits to these fixed spectral regions, and in (c) plot the estimated GDD as a function of current. The left lobe seems to have a GDD approximately twice that of the central region, while the right lobe has surprisingly a negative GDD of about the magnitude as the middle. As the current increases (i.e. the spectrum grows), the 3 values seem to all decrease in magnitude, though at no point do they reconcile.

	\section{Merit function}
	\label{Asec:merit}
	We refer interested readers to ~\cite{villares_quantum_2015} for a full discussion of the model.
	
	The merit function is calculated as follows. Firstly, the FWM term on the RHS of \ref{eq:merit} is computed for an amplitude spectrum of fixed magnitude, but swept chirp. In the second step, the overall merit is computed, including an adjustable parameter $\tilde{\alpha}$, which is set such that the difference between the LH terms and RH terms is minimised; the position of optimal chirp coincides with this term being maximised, the ultimate value of which can be interpreted as a measure of the efficiency of the lasing state.
	
	The modal amplitudes take on an assumed shape, in this particular case a sum of 3 Gaussians, but the total optical power is left as a free parameter by letting ${\tilde{A}_n=\alpha A_n}$.
	
	If we let
	\begin{equation}
	S_n = \sum_{k,l=-N/2}^{N/2} A_m A_k A_l^* B_{kl}C_{kl} \kappa_{n,l,k,m}
	\end{equation}
	be the FWM sum, where
	
	\begin{equation}
	C_{kl} = \frac{\gamma_{22}}{\gamma_{22}-i(k-l)\omega}
	\end{equation}
	
	\begin{equation}
	B_{kl} = \frac{\gamma_{12}}{2i}\left(\frac{1}{-i\gamma_{12}-l\omega}-\frac{1}{i\gamma_{12}-k\omega}\right)
	\end{equation}
	
	\begin{equation}
	\kappa_{n,l,k,m} = \begin{cases}
	\frac{3}{8}, & k=l=n,\\
	\frac{1}{4}, & k=l\ne n \text{ or } l=n, \\
	\frac{1}{8}, & \text{otherwise}
	\end{cases}
	\end{equation}
	
	we then have the following merit function:
	
	\begin{equation}
	M(\alpha,c) = \sum_n \left| (G_n -1) \tilde{\alpha}A_n - \tilde{\alpha}^3 G_n S_n \right|^2
	\label{eq:merit} 
	\end{equation}
	
	Where the objective is to minimise $M(\alpha,c)$.
	
\begin{figure}[!htb]
		\centering
		\includegraphics[width=\linewidth]{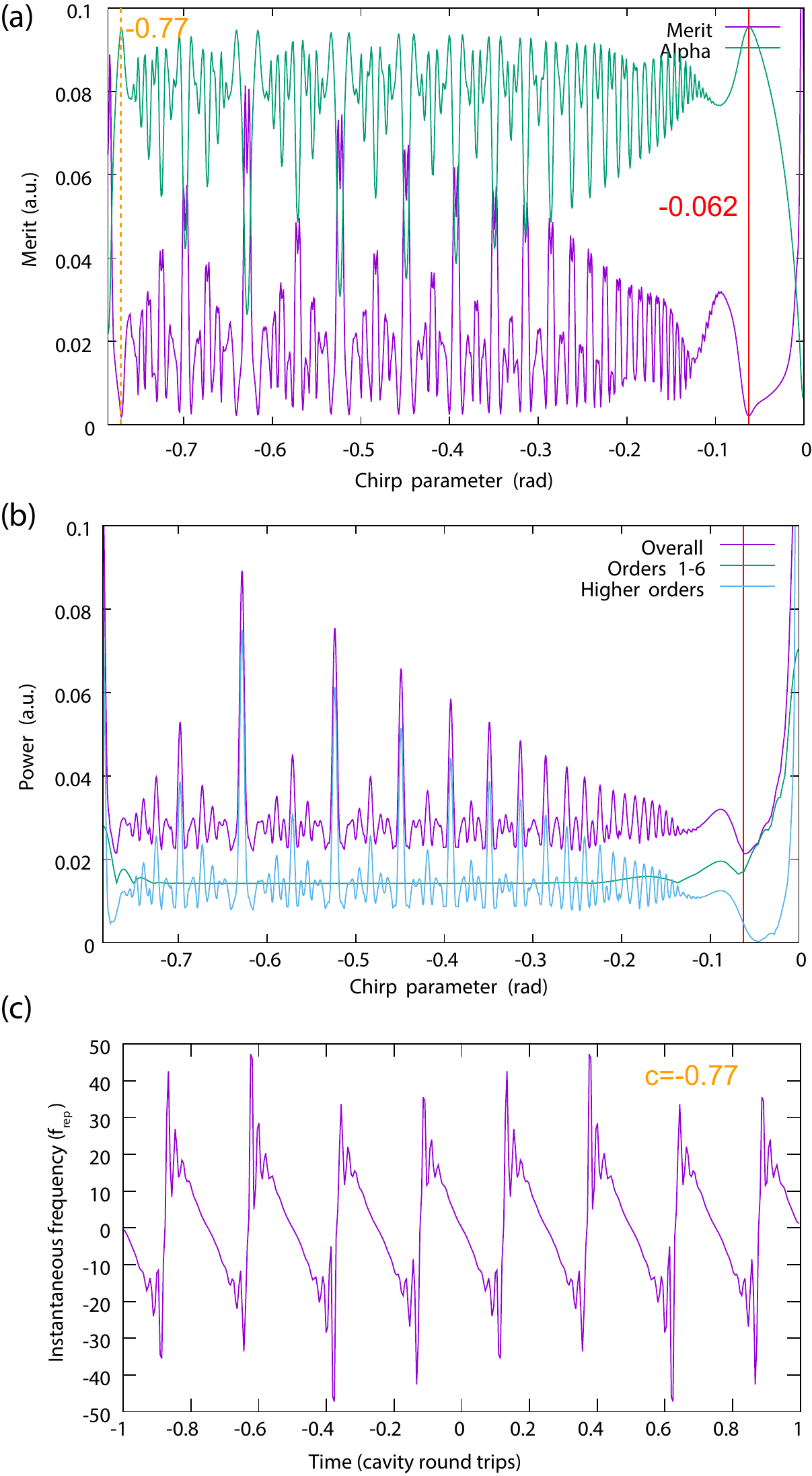}
		\caption{(a)Merit plotted for an extended range, showing the multiplicity of chirped solutions. Powers of low frequency (up to order 6) and high frequencies is superimposed, showing the sharp increase of high order beating terms, which we propose to induce losses. The red dashed line indicates a higher chirped state, for which the instantaneous frequency is plotted in (c). (b) Beating terms $\sum_n A_n A^*_{n+B}$, where $B$ is the beating order, computed as a function of the chirp parameter. The low and high orders of B are separated to highlight that high frequency beatings only play a role for higher chirped states.}
		\label{Afig:theory_merit}
	\end{figure}

	In Figure~\ref{Afig:theory_merit}~(a), we have plotted the merit function over a larger extent of chirp parameter, demonstrating that there are many values of chirp which minimise Eqn.~\ref{eq:merit}. One such state is depicted in (c), which makes 4 complete frequency sweeps per round-trip. 
	
	\section{Dissipation}
	\label{Asec:dissipation}
	As described by M. Piccardo, D. Kazakov et. al. \cite{piccardo_time-dependent_2018}, the optical field induces a spatiotemporal population grating in the device, externally measurable by probing across the length of the device the radio frequency voltage, oscillating at the intermodal beat frequencies. We propose that such a voltage grating would induce in-plane currents, which would contribute to the loss through ohmic dissipation.
	
	\begin{figure}[!htb]
		\centering
		\includegraphics[width=\linewidth]{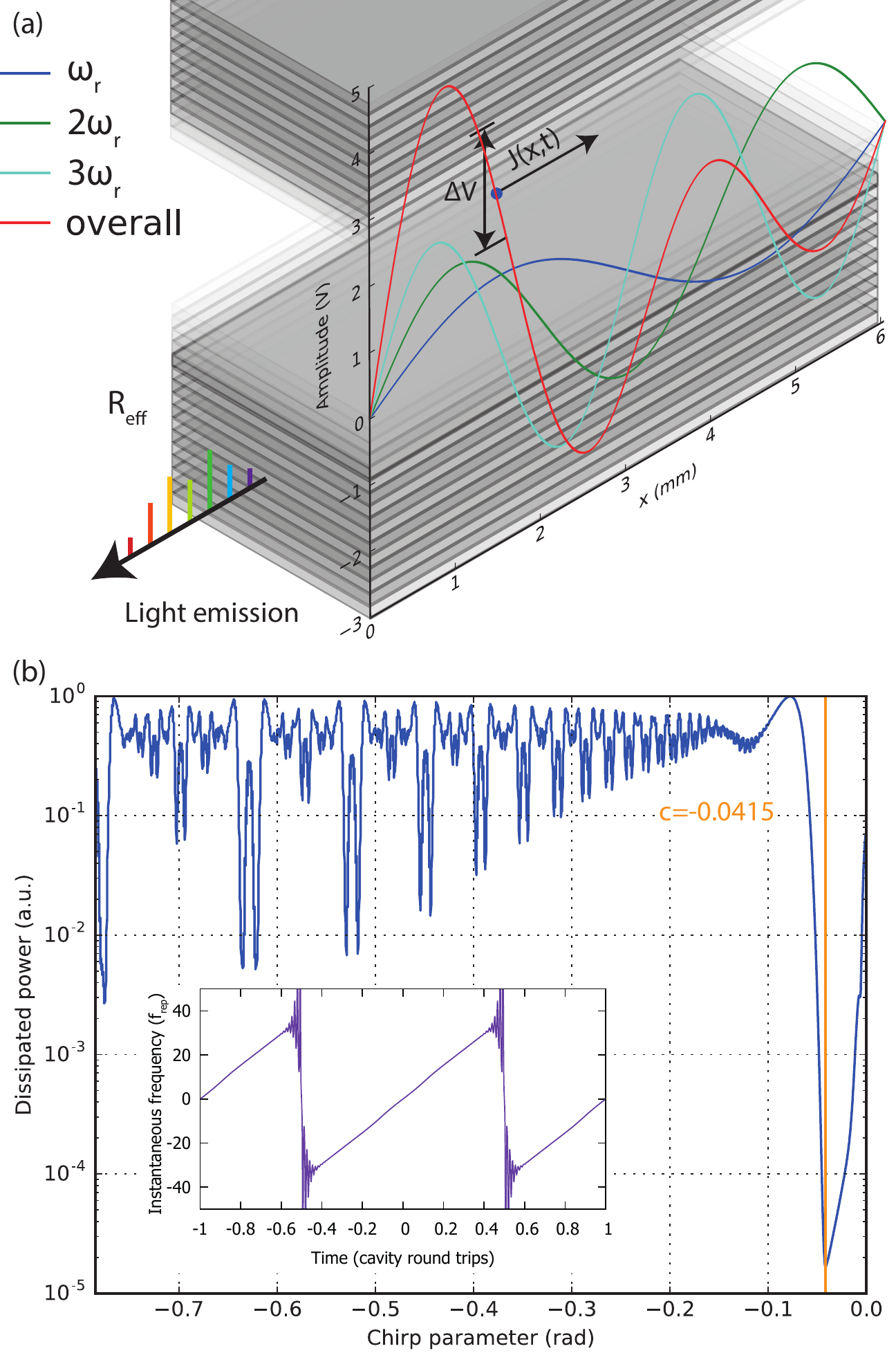}
		\caption{(a) QCL active region, with an intracavity field. Plotted are the intensity gratings at some fixed time $t_0$ for the fundamental and second and third harmonic of the intermodal beat frequency, and the sum of the 3. This is assumed to directly correspond to a voltage, which induces a current $\Delta V(x,t) R_eff$ (shown by the arrow) which varies spatiotemporally. (b) Dissipative loss, calculated for the same spectrum as in the main paper, as a function of chirp. This is scaled to $\alpha^2$, as in Figure~\ref{Afig:theory_merit}~(a).}
		\label{Afig:beating_dissipation}
	\end{figure}

	We begin by describing the intracavity field as a sum of forward and backward propagating waves ${E=E_f+E_b}$, assuming no mirror losses:
	\begin{equation}
	E_f = \sum_n A_n e^{-i\omega_n t} e^{k_n x}
	E_b = \sum_n A_n e^{i\omega_n t} e^{k_n x}
	\end{equation}
	
	The intensity is then $EE^*$ as usual:
	\begin{equation}
	I(x,t) = (E_f + E_b)(E_f+E_b)^* = E_f E_f^* + E_b E_b^* + E_f E_b^* + c.c.
	\end{equation}
	
	Then, noting that cross terms $E_f E_b^*$ produce oscillations on the order of $2\omega_n$, we neglect these. We have:
	
	\begin{equation}
	\begin{split}
	I(x,t) &= \sum_m \sum_n A_n A_m^* \Bigl [ \Bigr . \\
	 & \Bigl . e^{-i(n-m)\omega_r t} e^{i(k_n-k_m)x}+ e^{i(n-m)\omega_r t}e^{i(k_n-k_m)x} \Bigr]
	\end{split}
	\end{equation}
	
	If we then look at only those terms which oscillate at $b \omega_r$ (i.e. $m=n+b, m=n-b$):
	
	\begin{equation}
	\begin{split}
	\langle I(x,t) \rangle |_{b\omega_r} & = \sum_n A_n [ A_{n-b}^* e^{-i b \omega_r t} e^{i (k_n - k_{n-b})x} +\\
	& A_{n+b} e^{i b\omega_r t} e^{i (k_n-k_{n+b})x} +\\
	& A_{n-b}^* e^{i b \omega_r t} e^{i(k_n-k_{n-b})x}  +\\
	& A_{n+b}^* e^{-i b \omega_r t} e^{i(k_n-k_{n+b})x}  ]
	\end{split}
	\end{equation}
	
	Which, with the spatial repetition rate $k_r=2\pi(\frac{1}{\lambda_n}-\frac{1}{\lambda_{n-1}})$, conveniently simplifies to:
	
	\begin{equation}
	\langle I(x,t) \rangle |_{b\omega_r} = \cos(b\omega_r t) \sum_n A_n \left [
	A_{n-b}^* e^{i b k_r x} + A_{n+b}^* e^{-i b k_r x}
	\right ]
	\end{equation}
	
	\begin{equation}
	\langle I(x,t) \rangle |_{b\omega_r} = 
	\cos(b k_r x) \cos(b\omega_r t) \sum_n A_n A^*_{n-b} + c.c.
	\end{equation}
	
	The intra-cavity intensity will induce a population grating, which tracks these intensity beatings\cite{piccardo_time-dependent_2018}. As such will also exist  a spatiotemporally varying voltage, which we assume to run perfectly in phase with the driving intensity. This can therefore be expressed:
	
	\begin{equation}
	\label{eq:intensity_voltage}
	V(x,t)  = I R_{eff} \propto E E^*
	\end{equation}
	
	where $R_{eff}$ can be seen as an effective, or average, resistance, as experienced by the current passing through the structure in-plane. The situation is depicted in Figure~\ref{Afig:beating_dissipation}~(a), where the instantaneous potential difference $V(x_0+\delta,t)-V(x_0,t)$ is seen to induce a current density $J(x_0,t)$. We hence take the spatial derivative in the direction of propagation:
	
	\begin{equation}
	\frac{d V(x,t)}{d x}\Bigg |_{b \omega_r} = -b k_r \sin(b k_r x)\cos(b\omega_r t) + c.c.
	\end{equation}
	
	The power dissipated by order $b$ is therefore proportional to:
	\begin{equation}
	\label{eq:diss1}
	P_{diss}^{(b)} \propto b^2 K_r^2 \int{\int{\sin^2(b k_r x)\cos^2(b\omega_r t) d x} d t}
	\end{equation}
	
	i.e., the loss scales with the square of the beating frequency. Now, with reference to beating map shown in Figure~\ref{fig:theory}~(c), one might therefore expect these losses to increase after around $c=-0.05$, when the fast oscillations start to set in. By summing across all beats, we find the overall dissipation to be:
	
	\begin{equation}
	\begin{split}
	P_{diss} &= \frac{1}{T}\int_0^T \int_0^L \Bigl( \Bigr . \\
	        & \Bigl . -\sum_b \sin(b k_r x) \cos(b\omega_r t) \sum_n A_n A_{n-b}^* + c.c. \Bigr )^2  dx dt
	\end{split}
	\end{equation}
	
	In Figure\ref{Afig:beating_dissipation}~(b), we present this dissipation as a function of chirp. Under the assumptions we have made, one can see a clear penalty of a factor more than 100 with respect to the dissipation when lasing at a higher chirped state, which was not apparent from the merit function alone; indeed, the minimum is also slightly shifted towards a lower chirp of $c=-0.0415$. However, we should be careful in drawing conclusions from this, in that we have yet to quantify in real units the order of magnitude of this effect, and we have furthermore treated the active region as having a fixed effective resistance, rather than a frequency dependent impedance, which will no doubt again affect the weightings of the individual beating components. These need to be considered more carefully in a future work. Nonetheless, this idea of intracavity field intensity induced loss may go some way to explaining why the laser opts for a lower chirped state, where we have only one frequency sweep per round trip.

	\bibliographystyle{abbrv}
	\bibliography{biblio}
	

\end{document}